\documentclass[12pt]{article}
\usepackage{graphics,epsfig}
\usepackage[english]{babel}
\newcommand{\cinst}[2]{$^{\mathrm{#1}}$~#2\par}
\newcommand{\crefi}[1]{$^{\mathrm{#1}}$}

\begin{document}


\thispagestyle{empty}
\begingroup


\vglue.5cm \hspace{9cm}{YerPhI Preprint - 1625 (2012)
\vglue.5cm
\begin{center}

{\normalsize \bf STUDY OF THE $\Delta^{++}(1232)$ INCLUSIVE
\\[.3cm]
NEUTRINOPRODUCTION ON PROTONS AND NEUTRONS}

\end{center}

\vspace{1.cm}
\begin{center}

 N.M.~Agababyan\crefi{1},
 N.~Grigoryan\crefi{2}, H.~Gulkanyan\crefi{2},\\
 A.A.~Ivanilov\crefi{3}, V.A.~Korotkov\crefi{3}

\setlength{\parskip}{0mm} \small

\vspace{1.cm} \cinst{1}{Joint Institute for Nuclear Research,
Dubna, Russia} \cinst{2}{Alikhanyan National Scientific Laboratory \\
(Yerevan Physics Institute), Armenia}
\cinst{3}{Institute for High Energy Physics, Protvino, Russia}
\end{center}

\setlength{\parskip}{0mm}
\small

\vspace{130mm}

{\centerline{\bf YEREVAN  2012}}

\newpage
\vspace{1.cm}
\begin{abstract}
For the first time, the total yield and inclusive spectra of the
$\Delta^{++}(1232)$ isobar are measured in $\nu p$ and $\nu n$
charged-current interactions. An indication is obtained that the
$\Delta^{++}(1232)$ production mainly results from the neutrino
scattering on the valence $d$- quark of the target nucleon. The
total yield of $\Delta^{++}(1232)$ in $\nu p$ interactions is
compatible with that measured in hadronic interactions of the same
net charge and net baryonic number. The yield of
$\Delta^{++}(1232)$ in $\nu n$ interactions is significantly
suppressed as compared to the case of the proton target. The form
of the squared transverse momentum distributions, both in $\nu p$
and $\nu n$ interactions, is found to be compatible with the
available data on the neutrinoproduction of $\Lambda$ hyperon. The
experimental data are compared with the LEPTO6.5 model
predictions.
\end{abstract}

\newpage
\setcounter{page}{1}
\begin{center}
{\large 1. ~INTRODUCTION}\\
\end{center}

Hadron resonances play a prominent role in the multiparticle
production processes. In particular, they are important
ingredients in the fragmentation of leptoproduced quark strings.
The experimental data on the resonance production are necessary
for a better understanding of the hadron formation space-time
pattern. At present, more or less detailed experimental data on
the inclusive leptoproduction are collected for mesonic resonances
(see \cite{ref1} and references therein for the case of the
neutrinoproduction), while the data on the total and differential
yields of non-strange baryon resonances are obtained only for the
$\Delta^{++}(1236)$ isobar production in muon-proton scattering
\cite{ref2}.

The purpose of this paper is to obtain, for the first time, the
data on the total and partial yields and the differential spectra
of the $\Delta^{++}(1232)$ isobar in charged-current $\nu p$ and
$\nu n$ interactions. The experimental procedure is described in
Section 2. Section 3 presents the experimental results. Section 4 is devoted
to the comparison of the experimental data with the predictions of the LEPTO6.5
Monte-Carlo event generator. The results are summarized in Section 5.

\begin{center}
{\large 2. ~EXPERIMENTAL PROCEDURE}\\
\end{center}

The experiment was performed with SKAT bubble chamber \cite{ref3},
exposed to a wideband neutrino beam obtained with a 70 GeV primary
protons from the Serpukhov accelerator. The chamber was filled
with a propane-freon mixture containing 87 vol\% propane
($C_3H_8$) and 13 vol\% freon ($CF_3Br$) with the percentage of
nuclei H:C:F:Br = 67.9:26.8:4.0:1.3 \%. A 20 kG uniform magnetic
field was provided within the operating chamber volume.

Charged-current interactions containing a negative muon with
momentum $p_{\mu} >$0.5 GeV/c were selected. The overwhelming part of
protons with momentum below 0.6 GeV$/c$ and a fraction of protons  with momentum
0.6-0.85 GeV$/c$ were identified by their stopping in the chamber.
Stopping $\pi^+$ mesons were identified by their
$\pi^+$-$\mu^+$-$e^+$ decay. A fraction of low-momentum
($p_{\pi^+} < 0.5$ GeV$/c$) $\pi^+$ mesons were identified by the
mass-dependent fit provided that the $\chi^2$- value for the pion
hypothesis was significantly smaller as compared to that for
proton. Non-identified positively charged hadrons are assigned the pion mass or,
in the cases explained below, the proton mass.
It was required the errors in measuring the momenta be
less than 24\% for muon, 60\% for other charged particles and
$V^0$'s (corresponding to strange particles) and less than 100\%
for photons. The mean relative error ($\Delta p/p$) in the momentum measurement for
muons, pions, protons and gammas was, respectively, 3\%, 6.5\%, 10\% and 19\%.
Each event was given a weight to correct for the
fraction of events excluded due to improperly reconstruction. More
details concerning the experimental procedure, in particular, the
reconstruction of the neutrino energy $E_\nu$, can be found in our
previous publications \cite{ref4,ref5}.

The events with $3< E_{\nu} <$ 30 GeV were accepted, provided that
they satisfy the following topological and kinematic criteria for
quasinucleon interactions: the net charge of secondary hadrons was
equal to +1 (for quasineutron interactions) or +2 (for quasiproton
interactions); the number of recorded baryons (these included
identified protons and $\Lambda$ hyperons, along with neutrons
that suffered a secondary interaction in the chamber) was
forbidden to exceed unity, baryons flying in the backward
hemisphere being required to be absent among them; the effective
target mass $M_t <  $ 1.2 GeV/$c^2$, the $M_t$ being defined as
$M_t = \sum{(E_i - p_i^L)}$ where the summation was performed over
the energies $E_i$ and the longitudinal momenta $p_i^L$ (along the
neutrino direction) of all recorded secondary particles. A
detailed motivation of the aforementioned selection criteria can
be found in \cite{ref6} and references therein.

The numbers of accepted events corresponding to neutrino-proton
and neutrino-neutron interactions are equal to 1839 and 2393, with
the mean values of the neutrino energy $\langle E_\nu\rangle =8.9$
and 8.7 GeV, respectively. Only in a small fraction of these
events (23.7\% in $\nu p$ and 16.3\% in $\nu n$ interactions) an
identified proton ($n_p^{id} = 1$) was present. In the remaining
events (with $n_p^{id}$ = 0) the proton hypothesis was applied to
a non-identified positively charged hadron (if any)  provided that
the proton hypothesis was not rejected by the momentum-range
relation in the propane-freon mixture. This hypothetical proton,
after introduction of a proper correction for its momentum, was
combined with an accompanying positively charged hadron (an
identified $\pi^+$ or a non-identified hadron) to compose a
hypothetical $\pi^+ p$ combination. The most part of such
combinations, especially in events containing two or more
unidentified hadrons ($n_h^{nid} \geq 2$), is expected to be
spurious due to the proton misidentification. As a result, the
angular distribution of a 'proton' in the pion-'proton' rest frame
turns out to be strongly shifted towards the negative values of
$\cos\vartheta^*_p$, where $\vartheta^*_p$ is the angle between
the 'proton' direction and the direction of the Lorentz boost from
the lab system to the pion-'proton' rest system. As it has been
shown by simulations based on the LERTO6.5 Monte-Carlo
generator\cite{ref7}, the  $\cos\vartheta^*_p$ distribution for
the case of spurious $\pi^+ p$ combinations is strongly peaked at
$\cos\vartheta^*_p \approx -1$ and rapidly falls with increasing
$\cos\vartheta^*_p$ up to $\cos\vartheta^*_p \sim -0.6$, then
begins flatten and becomes almost uniform at $\cos\vartheta^*_p
>0$. In order to reduce the share of spurious combinations in the
experimental $\pi^+ p$ effective mass distribution for events with
$n_p^{id} = 0$ and $n_h^{nid} \geq 2$, a cut $\cos\vartheta^*_p >
-0.6$ was applied for the combinations of two unidentified
hadrons, while those with $-0.6 < \cos\vartheta^*_p < 0$ a weight
were ascribed, so that the total numbers of combinations with
$\cos\vartheta^*_p < 0$ and $\cos\vartheta^*_p > 0$ (including
those for events with $n_p^{id} = 1$ or $n_h^{nid} < 2$) turned
out to be equal in the $\Delta^{++}(1236)$ peak region of $1.16 <
m_{\pi^+ p} < 1.32$ GeV$/c^2$. This procedure, as it will be
discussed in the next section, enables to improve the signal to
background ratio in the $\Delta^{++}(1236)$ peak region and
somewhat decreases the errors in the determination of its yield.

\begin{center}
{\large 3. ~EXPERIMENTAL RESULTS}\\
\end{center}

The $\pi^+ p$ effective mass distributions for $\nu p$ and $\nu n$
interactions are plotted in Figure 1. They can be described by the
sum of a Breit-Wigner function \cite{ref8}, smeared according to
the experimental resolution, and a background distribution $BG(m)$
parametrized as (following \cite{ref2})

\begin{center}
$BG(m) = B \cdot q^\alpha \cdot \exp(-\beta m^\alpha)$ \, ,
\end{center}

\noindent where $B, \alpha, \beta$ and $\gamma$ are free
parameters and $q$ is the pion momentum in the $\pi^+ p$ rest
frame.

The fit results in the following mean yields of
$\Delta^{++}(1232)$: $\langle n\rangle_{\nu p} = 0.181\pm0.025$
and \, \, $\langle n\rangle_{\nu n} = 0.051\pm0.012$. The quoted
errors here and below reflect also the uncertainty related to the
choice of initial values of the fit parameters. We also analyzed
our data without any elimination of combinations of two
unidentified hadrons and observed a worsening of the signal to
background ratio resulting in a smaller values and larger relative
errors of $\langle n\rangle_{\nu p}$ and $\langle n\rangle_{\nu
n}$ which turned out to be $0.164\pm0.030$ and $0.029\pm0.018$,
respectively. We have also tested the consistency of our data with
the partial yield of $\Delta^{++}(1232)$ in $\nu p$ interactions
measured in \cite{ref9} where only $\pi^+ p$ combinations with an
identified proton with momentum less than 1 GeV$/c$ were used. In
our experiment this partial yield turns out to be $\langle n(p_p <
1$ GeV$/c$)$\rangle_{\nu p}$ = 0.103$\pm$0.012 when a restriction
on $\cos \vartheta^*_p$ (described in the previous section) was
applied and $\langle n(p_p < 1$ GeV$/c$)$\rangle_{\nu p}$ =
0.095$\pm$0.015 when not. These values are quite compatible with
the value of 0.105$\pm$0.017 obtained in \cite{ref9}.

Table 1 presents the mean yields in
different ranges of the neutrino energy and the Bjorken $x_B$
variable. The quoted yields are normalized to the number of events
of the corresponding subsample.

\begin{table}[ht]
\caption{The dependence of $\langle n\rangle_{\nu p}$ and $\langle n\rangle_{\nu n}$ on
$E_\nu$ and $x_B$.}
\begin{center}
\begin{tabular}{|l|c|c|}
  \hline

The range of variable&$\langle n\rangle_{\nu p}$&$\langle n\rangle_{\nu n}$
\\ \hline
$3<E_\nu<7$ GeV&0.208$\pm$0.036&0.013$\pm$0.017 \\
$7<E_\nu<30$ GeV&0.179$\pm$0.037&0.072$\pm$0.027
\\ \hline
$x_B<0.2$&0.153$\pm$0.040&0.025$\pm$0.024
\\
$x_B>0.2$&0.227$\pm$0.035&0.052$\pm$0.018
\\ \hline

\end{tabular}
\end{center}
\end{table}

As it is seen, the yield of $\Delta^{++}(1232)$ at $x_B > 0.2$
exceeds that at $x_B < 0.2$, indicating that in its production the
dominant role belongs to the neutrino scattering on the valence
quark (in this case -- on the valence $d$- quark) of the target
nucleon. This feature, being less expressed, was observed earlier
in deep-inelastic $\mu p$ scattering \cite{ref2}. The yield
$\langle n\rangle_{\nu n}$ exhibits a strong dependence on $E_\nu$ (in
particular, no $\Delta^{++}$ production is observed in the
low-energy range of $3 < E_\nu <7 $ GeV). On the contrary, the
yield $\langle n\rangle _{\nu p}$ is almost independent of $E_\nu$. As it is seen
from Figure 2, the yield $\langle n\rangle _{\nu p}$ is within error compatible
with that measured in hadronic interactions with the same net
charge (+2) and baryonic number (+1) in the final state (see
\cite{ref10,ref11} and references therein).

We also attempted to measure the inclusive spectra of
$\Delta^{++}(1232)$ on its kinematic variables: the Feynman
$x_F$ variable, the squared transverse momentum $p_T^2$ (where
$p_T$ is defined respective to the intermediate boson direction),
and the $z$ variable -- the fraction of the exchanged boson energy
transferred to the $\Delta^{++}(1232)$ and defined as $z =
(E-m_p)/\nu$, where $E$ is the $\Delta^{++}$ total energy, $m_p$
is the proton mass and $\nu$ is the exchanged boson energy $\nu =
E_\nu - E_{\mu^-}$. Since these variables are senseless for the
exclusive reaction $\nu p \rightarrow \mu^- p
\pi^+$ (being fixed by definition: $x_F = 0$, $p_T = 0$, $z = 1$),
the events-candidates to the latter were below excluded from
consideration (see \cite{ref12} for details).

Figure 3 shows the $\pi^+ p$ effective mass distributions for the
full range of $x_F$, as well as for the $\pi^+ p$ system produced
in the backward $(x_F < 0)$ and forward $(x_F > 0)$ hemispheres of
the neutrinoproduced hadronic system. The corresponding mean
yields of $\Delta^{++}(1232)$ in $\nu p$ and $\nu n$ interactions
are presented in Table 2, together with the data obtained in $\mu
p$ interactions \cite{ref2}.

\noindent
\begin{table}[ht]
\caption{The $\Delta^{++}(1232)$ mean yields in different ranges
of $x_F$ in $\nu p$, $\nu n$ and $\mu p$ interactions.}
\begin{center}
\begin{tabular}{|l|c|c|c|}
  \hline

&$\nu p$&$\nu n$&$\mu p$ [2]
\\ \hline
all $x_F$& 0.170$\pm$0.029&0.051$\pm$0.012&0.10$\pm$0.02 \\
$x_F < 0$& 0.101$\pm$0.023&0.043$\pm$0.011&0.08$\pm$0.02 \\
$x_F > 0$& 0.057$\pm$0.018&0.010$\pm$0.009&0.02$\pm$0.02 \\ \hline

\end{tabular}
\end{center}
\end{table}

As it is seen, the yield of $\Delta^{++}(1232)$ for the case of
the neutron target is approximately threefold suppressed as compared
to the case of the proton target. This suppression is yet more
expressed in the forward hemisphere ($x_F > 0$). It is interesting
to note, that the magnitude of the yield of $\Delta^{++}(1232)$ in
$\mu p$ interactions occupies an intermediate position between
those in $\nu p$ and $\nu n$ interactions. The forward-backward
asymmetry parameter, defined as $A = (\langle n(x_F > 0)\rangle - \langle n(x_F < 0)\rangle)
/(\langle n(x_F
> 0)\rangle + \langle n(x_F < 0)\rangle)$, is equal to $A(\nu p) = -0.28\pm0.18$
and $A(\nu n) = -0.62\pm0.29$ for $\nu p$ and $\nu n$
interactions, respectively. These values can be compared with
$A(\mu p) = -0.60\pm0.33$ in $\mu p$ interactions \cite{ref2}, as
well as with the value of $A_\Lambda = -0.589\pm0.004$ measured in
the neutrinoproduction of $\Lambda$ hyperons at higher energies,
$\langle E_\nu\rangle = 45.3$ GeV \cite{ref13}.

The differential spectra of the $\Delta^{++}(1232)$ inclusive
production versus variables $z$, $x_F$ and $p_T^2$ are plotted in
Figures 4 and 5. As it is seen from Figure 4, the $z$-
distribution for $\nu p$ interactions is rather flat, while the
yield of the leading $\Delta^{++}(1232)$ (with $z >$ 0.5) in $\nu
n$ interactions is strongly suppressed. A similar suppression is
also observed at $x_F > 0$ in the $x_F$- distribution (Figure 5,
the left panel), both for $\nu n$ and $\mu p$ interactions. On the
other hand, the general trends of the $x_F$ distributions in $\nu
p$ and $\mu p$ interactions are almost the same in the backward
hemisphere $(x_F < 0)$. The $p_T^2$ distributions, both for $\nu
p$ and $\nu n$ interactions, can be described by an exponential
function with compatible slope parameters $b(\nu p) = 4.4\pm0.8$
and $b(\nu n) = 4.4\pm0.9$ (GeV$/c)^{-2}$, respectively (see the
solid and dashed lines in Figure 5, the right panel). These values
are compatible with those obtained for the neutrinoproduction of
$\Lambda$ (see \cite{ref13,ref14} and references therein). On the
contrary, the data for $\mu p$ interactions \cite{ref2} (also
plotted in Figure 5) can be satisfactorily described by a sum of
two exponential functions with the slope parameters $b_1(\mu p) =
11.6\pm5.7$ and $b_2(\mu p) = 1.9\pm0.7$ (GeV$/c)^{-2}$ (see the
dotted curve in Figure 5). As it was
shown in \cite{ref2}, the high-momentum tail of the $p_T^2$
distribution can be described in the framework of the Lund string
model \cite{ref15} modified according to the QCD effects. The role
of the latter in the high-$p_T$ region increases with increasing
$W$ (see \cite{ref16}, as well as \cite{ref17} and references
therein) and is expected to be more prominent in the high-$W$
region ($4 < W < 20$ GeV) at which the data \cite{ref2} on the
$\Delta^{++}(1232)$ muonoproduction are obtained.

\begin{center}
{\large 4. ~COMPARISON WITH LEPTO6.5 MODEL PREDICTIONS}\\
\end{center}

In this section we compare our data on the 
$\Delta^{++}(1232)$ yield in $\nu p$ and $\nu n$ interactions with the
predictions of the LERTO6.5 Monte-Carlo event generator for deep
inelastic lepton-nucleon scattering \cite{ref7} . Since the model puts a
restriction $W > 2$ GeV on the invariant mass of the leptoproduced
hadron system, the same should be done in the analysis of the
experimental data. Figure 6 shows the $\pi^+ p$ effective mass
distributions for $\nu p$ and $\nu n$ interactions at $W > 2$ GeV.
The corresponding inferred total yields, as well as those at restricted 
$x_B$ - ranges and the differential yields at $x_F < 0$ and $x_F > 0$ are 
quoted in Tables 3 and 4, along with the data on $\mu p$ interactions  
\cite{ref2} and the LERTO6.5 predictions at
the default values of the model parameters (except those related
to the experimental restrictions on the kinematic variables). 

As it is seen, the model predictions are within error compatible with the
data for $\nu n$ and $\mu p$ interactions. It should be pointed out, that 
the $\Delta^{++}(1232)$ production in $\nu n$ interactions at our 
energies is practically possible only owing to the diquark splitting in the 
target neutron remnant. The 'popcorn' \cite{ref18} scheme of this splitting incorporated 
into the model qualitatively reproduces our data on $\nu n$ interactions presented 
in Tables 3 and 4.  

\begin{table}[ht]
\caption{The $\Delta^{++}(1232)$ total yields for three ranges of the 
variable $x_B$ in $\nu p$ and $\nu
n$ interactions at $2 < W < 7$ GeV and in $\mu p$ interactions at
$4 < W < 20$ GeV compared to the LEPTO6.5 model predictions.}
\begin{center}
\begin{tabular}{|l|c|c|}
  \hline

Interaction&Experiment&Prediction \\ \hline
$\nu p$ && \\ 
all $x_B$ & 0.133$\pm$0.032&0.417 \\
$x_B < 0.2$&0.090$\pm$0.041&0.391  \\
$x_B > 0.2$&0.150$\pm$0.044&0.416  \\ \hline
$\nu n$ &&  \\
all $x_B$&0.068$\pm$0.026&$0.080$   \\
$x_B < 0.2$&0.036$\pm$0.028&0.073  \\
$x_B > 0.2$&0.098$\pm$0.039&0.082  \\ \hline
$ \mu p$ &&   \\ 
all $x_B$ &0.10$\pm$0.02&0.108     \\  
$x_B < 0.1$&0.09$\pm$0.02&0.125  \\
$0.1 < x_B < 0.2$&0.10$\pm$0.02&0.115  \\
$x_B > 0.2$&0.13$\pm$0.03&0.091  \\ \hline

\end{tabular}
\end{center}
\end{table}

As for the $\Delta^{++}(1232)$ production on the proton target, the model 
predictions badly overestimate (by about three times) the total and differential yields quoted in Tables 3 and 4. We failed to reach a satisfactory and simultaneous 
description of the experimental data presented in Tables 3 and 4 by a unique set of the model input parameters.

\begin{table}[ht]
\caption{The $\Delta^{++}(1232)$ differential yields 
at $x_F < 0$ and $x_F > 0$ in $\nu p$ and $\nu
n$ interactions at $2 < W < 7$ GeV and in $\mu p$ interactions at
$4 < W < 20$ GeV compared to the LEPTO6.5 model predictions.}
\begin{center}
\begin{tabular}{|l|c|c|}
  \hline

Interaction&Experiment&Prediction \\ \hline
$\nu p$ && \\ 
$x_F < 0$&0.084$\pm$0.025&0.240  \\
$x_F > 0$&0.054$\pm$0.017&0.177  \\ \hline
$\nu n$ &&  \\
$x_F < 0$&0.056$\pm$0.018&0.041  \\
$x_F > 0$&0.016$\pm$0.014&0.039  \\ \hline
$ \mu p$ &&   \\ 
$x_F < 0$&0.08$\pm$0.02&0.089  \\
$x_F > 0$&0.02$\pm$0.02&0.019  \\ \hline

\end{tabular}
\end{center}
\end{table}

\newpage

\begin{center}
{\large 5. ~SUMMARY}\\
\end{center}

First experimental data on the total yield and inclusive spectra
of $\Delta^{++}(1232)$ in $\nu p$ and $\nu n$ interactions are
obtained. The total yield of $\Delta^{++}(1232)$ in $\nu p$
interactions is compatible with that measured in hadronic
interactions of the same net charge and net baryonic number. The
yield of $\Delta^{++}(1232)$ at $x_B > 0.2$ exceeds noticeably
that at $x_B < 0.2$, indicating that the neutrino scattering on
the valence $d$- quark of the target proton or neutron plays a
dominant role in the $\Delta^{++}(1232)$ production process. The
$\Delta^{++}(1232)$ yield for the case of the neutron target is
significantly suppressed as compared to the case of the proton
target. This suppression is more expressed in the forward
hemisphere $(x_F
> 0)$, as well as for the leading $\Delta^{++}(1232)$ with $z >
0.5$. The $p_T^2$ distributions can be described by an exponential
function with slope parameters $b(\nu p)$ = 4.4$\pm$0.8 and $b(\nu
n) = 4.4\pm0.9$ (GeV$/c)^{-2}$ which are compatible with those for
$\Lambda$ hyperon in neutrino-induced reactions. The LEPTO6.5
model predictions agree with the measured $\Delta^{++}(1232)$
total yield at $W > 2$ GeV for $\nu n$ interactions, but
overestimate by about 3 times that for $\nu p$ interactions.

\begin{center}
{\large ACKNOWLEDGMENTS}\\
\end{center}

The authors from YerPhI acknowledge the supporting grants of
Calouste Gulbenkian Foundation and Swiss Fonds Kidagan. The
activity of one of the authors (H.G.) is supported by Cooperation
Agreement between DESY and YerPhI signed on December 6, 2002.
One of the authors (H.G.) is grateful to A. Kotzinian for helpful
discussions.

\newpage
\begin{figure}[ht]
\resizebox{0.9\textwidth}{!}{\includegraphics*[bb =20 65 600
610]{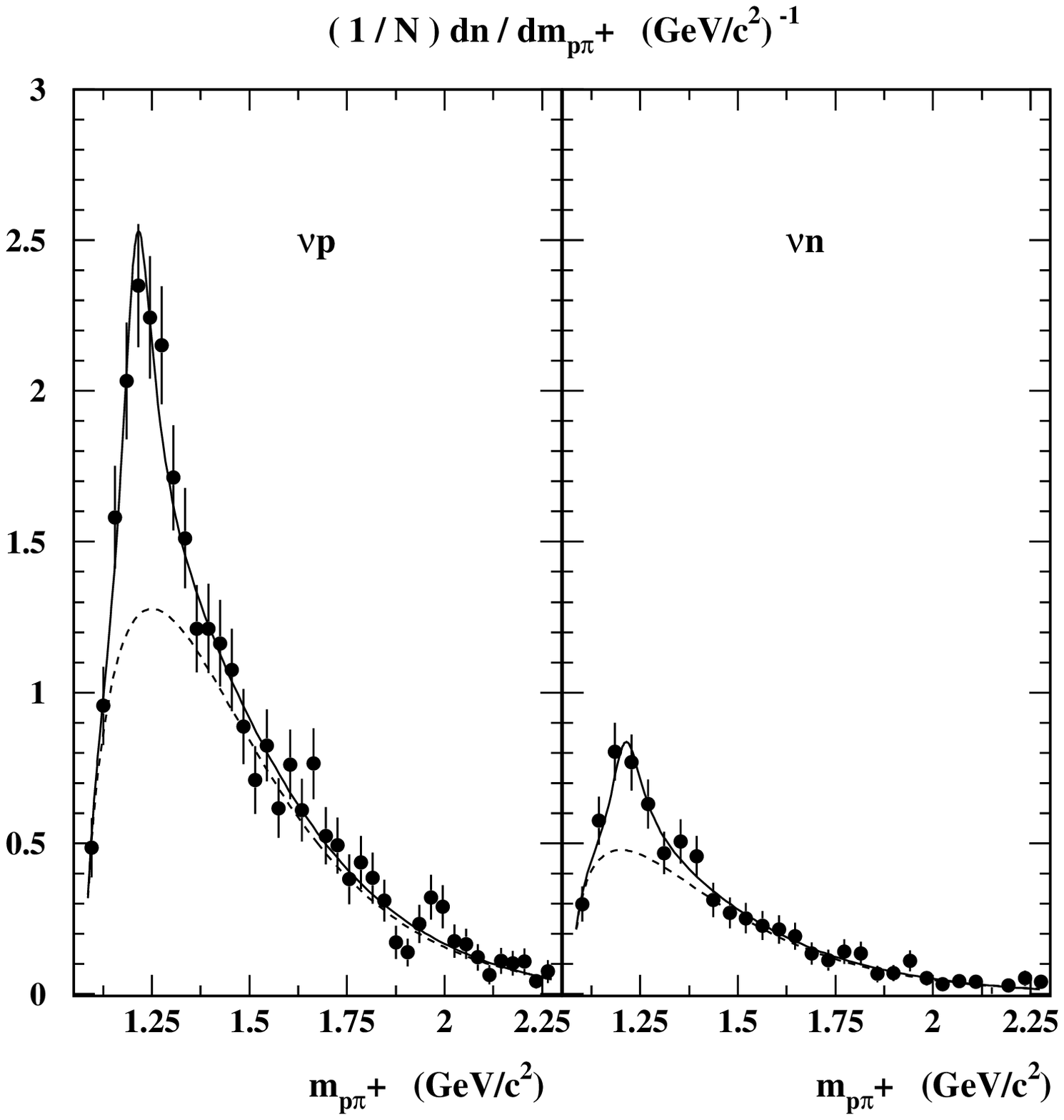}} \caption{The $\pi^+ p$ effective mass
distributions in $\nu p$ and $\nu n$ interactions. The curves are
the fit result (the background distribution is shown by dashed
curves).}
\end{figure}

\newpage
\begin{figure}[ht]
\resizebox{0.9 \textwidth}{!}{\includegraphics*[bb=20 40 500 610]
{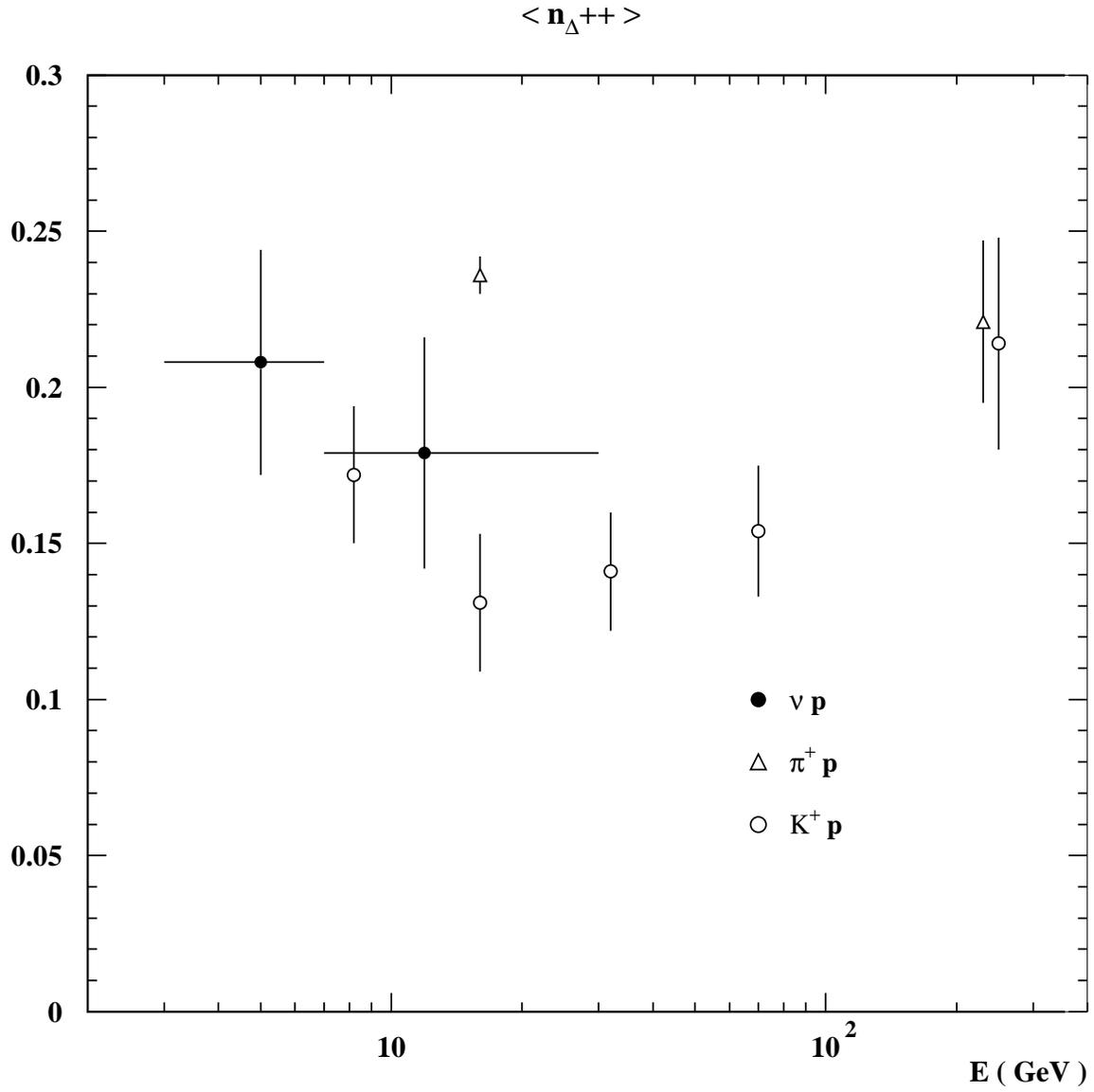}} \caption{The incident energy dependence of the
$\Delta^{++}(1232)$ mean yield in $\nu p$, $\pi^+ p$ and $K^+ p$
interactions.}
\end{figure}

\newpage
\begin{figure}[ht]
\resizebox{0.9\textwidth}{!}{\includegraphics*[bb =20 65 600
610]{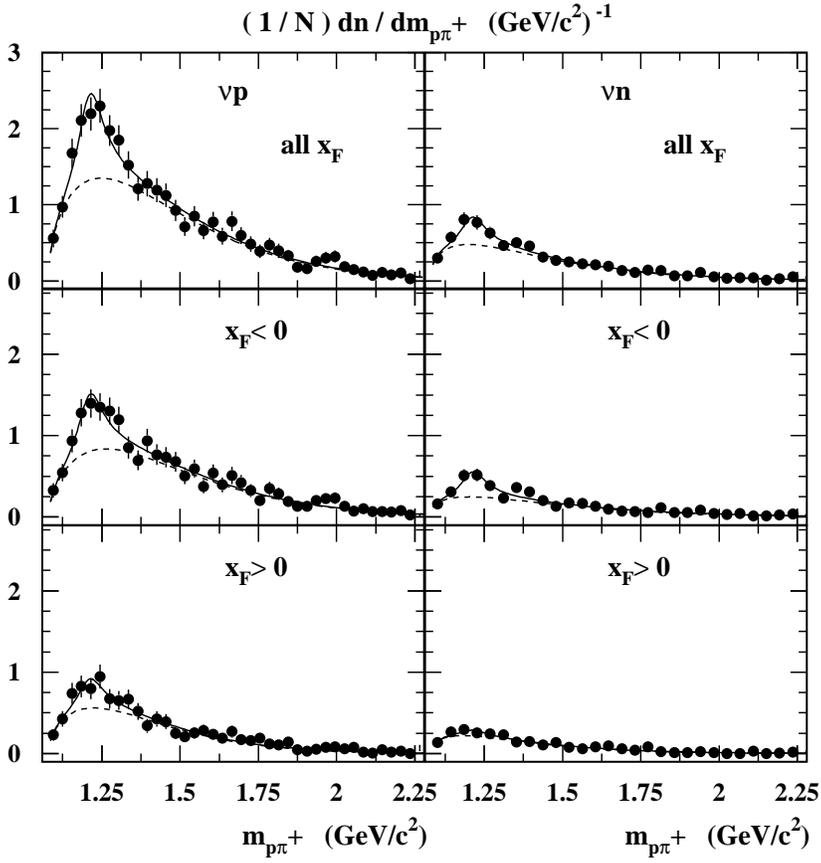}} \caption{The $\pi^+ p$ effective mass
distributions in $\nu p$ and $\nu n$ interactions for three
different ranges of $x_F$. The events-candidates to the reaction
$\nu p \rightarrow \mu^- p \pi^+$ are excluded. The meaning of
curves is the same as for Figure 1.}
\end{figure}

\newpage
\begin{figure}[ht]
\resizebox{0.9\textwidth}{!}{\includegraphics*[bb =20 65 600
610]{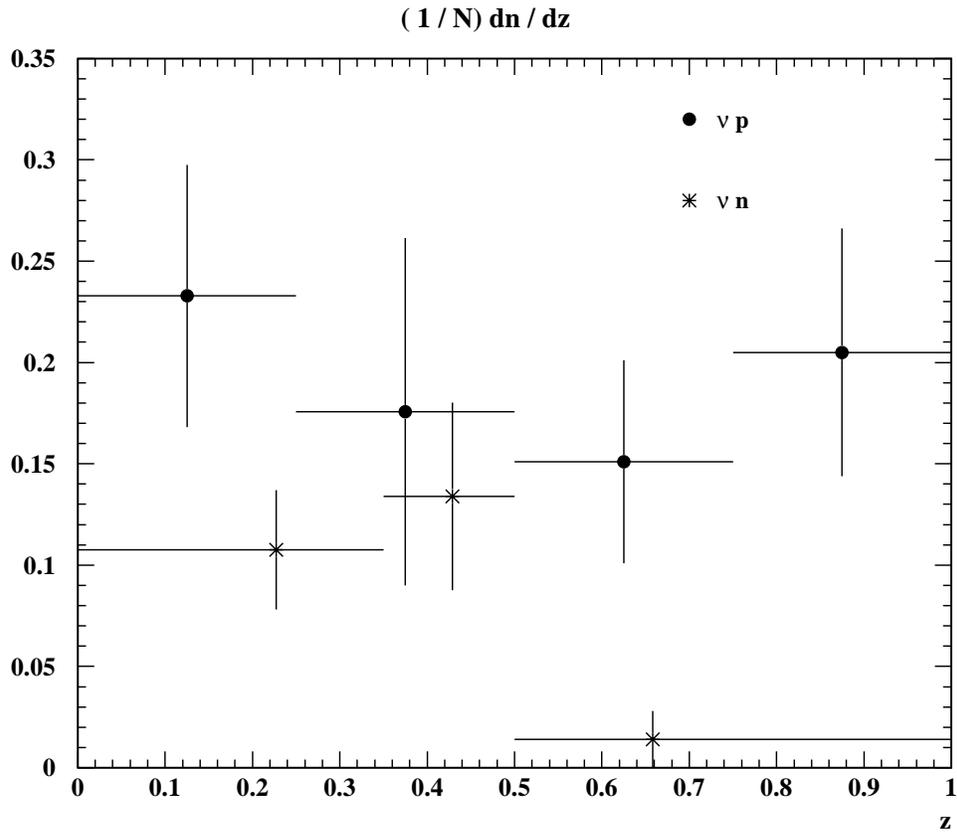}} \caption{The differential spectra of
$\Delta^{++}(1232)$ on the $z$ variable in $\nu p$ and $\nu n$
interactions.}
\end{figure}

\newpage
\begin{figure}[ht]
\resizebox{0.9\textwidth}{!}{\includegraphics*[bb =20 65 600 610]{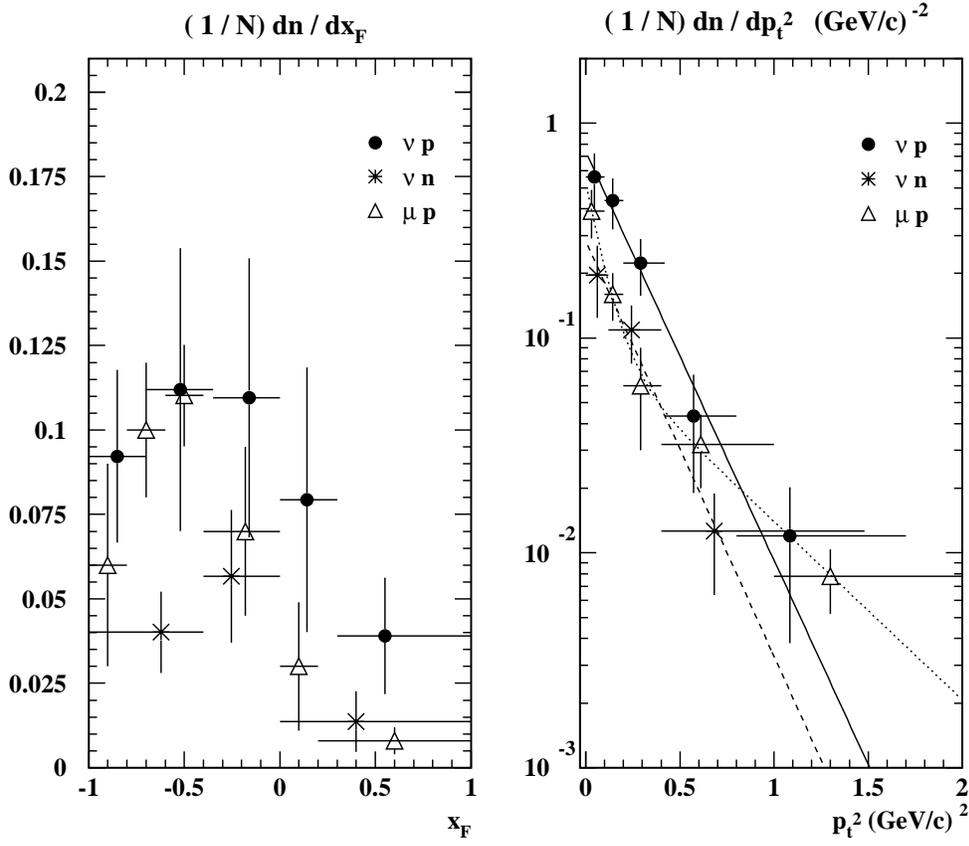}}
\caption{The differential spectra of
$\Delta^{++}(1232)$ on $x_F$ and $p_T^2$ variables in $\nu p$,
$\nu n$ and $\mu p$ interactions. The curves are the fit result
(see the text).}
\end{figure}

\newpage
\newpage
\begin{figure}[ht]
\resizebox{0.9\textwidth}{!}{\includegraphics*[bb =20 65 600 610]{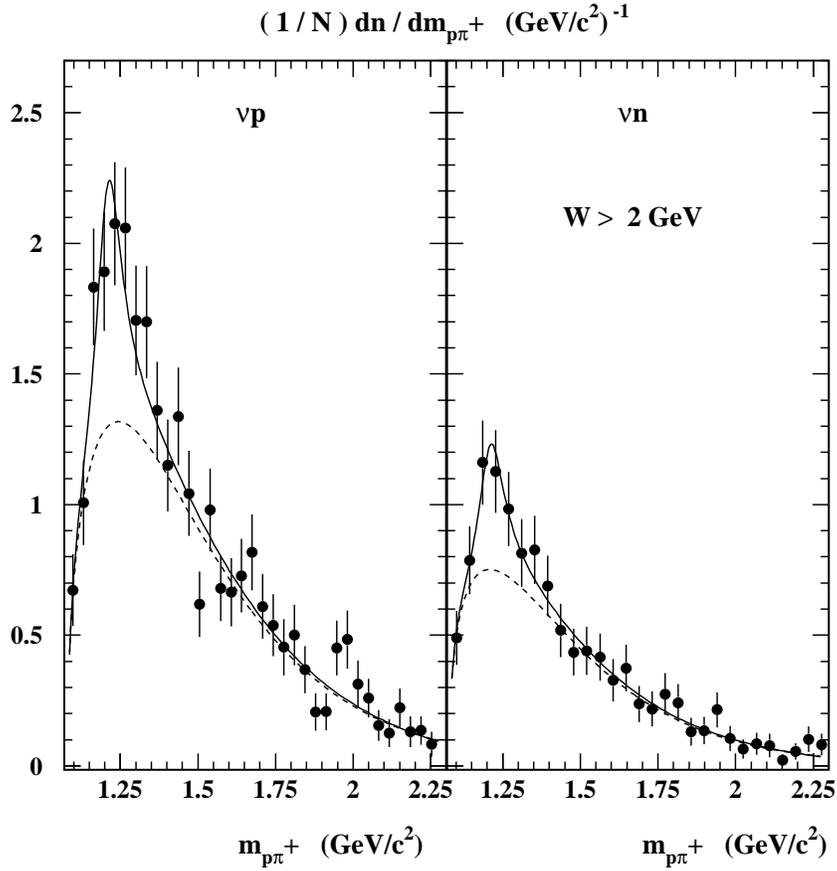}}
\caption{The $\pi^+ p$ effective mass distributions in $\nu p$ and
$\nu n$ interactions at $W > 2$ GeV. The meaning of curves is the
same as for Figure 1.}

\end{figure}
\end{document}